\documentclass[conference,10pt]{IEEEtran}
\IEEEoverridecommandlockouts
\usepackage{cite}
\usepackage{amsmath,amssymb,amsfonts}
\usepackage{algorithmic}
\usepackage{graphicx}
\usepackage{textcomp}
\usepackage{xcolor}
\usepackage{subfig}
\usepackage{bm}

\def\BibTeX{{\rm B\kern-.05em{\sc i\kern-.025em b}\kern-.08em
    T\kern-.1667em\lower.7ex\hbox{E}\kern-.125emX}}
\graphicspath{{figures/drawings/}{figures/plots/pdf/}}

\begin{document}

\title{Deep Learning Approaches for Open Set Wireless Transmitter Authorization}

\author{\IEEEauthorblockN{Samer Hanna, Samurdhi Karunaratne, and Danijela Cabric 
\thanks{This work was supported in part by the CONIX Research Center, one of six centers in JUMP, a Semiconductor Research Corporation (SRC) program sponsored by DARPA.}
		}
\IEEEauthorblockA{\textit{Electrical and Computer Engineering Department,} \\
\textit{University of California, Los Angeles}\\
samerhanna@ucla.edu, samurdhi@ucla.edu, danijela@ee.ucla.edu }
}

\IEEEoverridecommandlockouts
\IEEEpubid{\makebox[\columnwidth]{\copyright2020 IEEE \hfill} \hspace{\columnsep}\makebox[\columnwidth]{ }}
\IEEEpubidadjcol

\maketitle

\begin{abstract}
Wireless signals contain transmitter specific features, which can be used to verify the identity of transmitters and assist in implementing an authentication and authorization system. Most recently, there has been a wide interest in using deep learning for transmitter identification. However, the existing deep learning work has posed the problem as closed set classification, where a neural network classifies among a finite set of known transmitters. No matter how large this set is, it will not include all transmitters that exist. Malicious transmitters outside this closed set, once within communications range, can jeopardize the system security. In this paper, we propose a deep learning approach for transmitter authorization based on open set recognition. Our proposed approach identifies a set of authorized transmitters, while rejecting any other unseen transmitters by recognizing their signals as outliers. We propose three approaches for this problem and show their ability to reject signals from unauthorized transmitters on a dataset of WiFi captures. We consider the structure of training data needed, and we show that the accuracy improves by having signals from known unauthorized transmitters in the training set.

\end{abstract}


\section{Introduction}
With the growth in the number of wirelessly connected devices, securing  them has become more challenging. Part of securing wireless devices is authentication; the process of verifying their identity. While there exist many cryptography based methods for authentication, they are not suitable for many internet-of-things devices that have limited computation and power budget.

Physical layer authentication (PLA) enables devices to be authenticated without having to decode the data and typically without  requiring additional signaling overhead~\cite{yu_physical-layer_2008}. Active PLA overlays a tag for authentication over the message thus requiring changes to the physical layer of the transmitters. Passive PLA, on the other hand, uses the channel state information and the transmitter fingerprint due to hardware imperfections to identify transmitters ~\cite{wang_wireless_2016}, requiring no change to transmitter signals, and hence is more practical.

Approaches for passive PLA either use a set of handcrafted features  or deep learning on raw IQ samples.
For feature-based PLA, existing works have considered using  transmitter fingerprints due to hardware imperfections~\cite{peng_design_2019} or  channel state information (CSI)~\cite{xiao_using_2008}. Learning approaches based on handcrafted features rejecting new transmitters have used Gaussian mixture models~\cite{xu_device_2016-1,nguyen_device_2011,xiao_using_2008,gulati_gmm_2013}.  However, the performance of these approaches depends on the receiver quality \cite{rehman_analysis_2012} and requires manual feature engineering.

 In contrast, deep learning approaches are more robust and can extract better features from  signals, hence leading to higher accuracy compared to feature-based approaches \cite{riyaz_deep_2018}. %
   The existing work in the literature has considered the effect of data representation, neural network architecture, and the wireless channel on the classification accuracy~\cite{yu_robust_2019,gopalakrishnan_robust_2019,baldini_comparison_2019,agadakos_deep_2019,wu_deep_2018,riyaz_deep_2018,merchant_deep_2018-1,hanna_icnc_2019,youssef_machine_2017}. Examples of data representations include raw IQ samples \cite{riyaz_deep_2018,merchant_deep_2018-1,gopalakrishnan_robust_2019,yu_robust_2019}, Fourier transform \cite{hanna_icnc_2019,baldini_comparison_2019}, and  Wavelet transform \cite{youssef_machine_2017,baldini_comparison_2019}.  The  robustness of the learned features in different channels has also been considered \cite{yu_robust_2019}. Network architectures evaluated include DNN \cite{youssef_machine_2017}, CNN \cite{hanna_icnc_2019,riyaz_deep_2018,youssef_machine_2017}, RNN \cite{wu_deep_2018}, and complex neural networks~\cite{agadakos_deep_2019}. 
  The main limitation of this body of work is its focus on classification among a closed set of known transmitters. Any transmitter outside this set will be misclassified, hence, jeopardizing the system security.  The problem of classifying among known classes and  rejecting samples from new classes is known as open set recognition~\cite{openset_survey_2019}. Many approaches have been proposed to address it in other domains like image classification.

\begin{figure}[t!]
	\centering
	\includegraphics[scale=0.7]{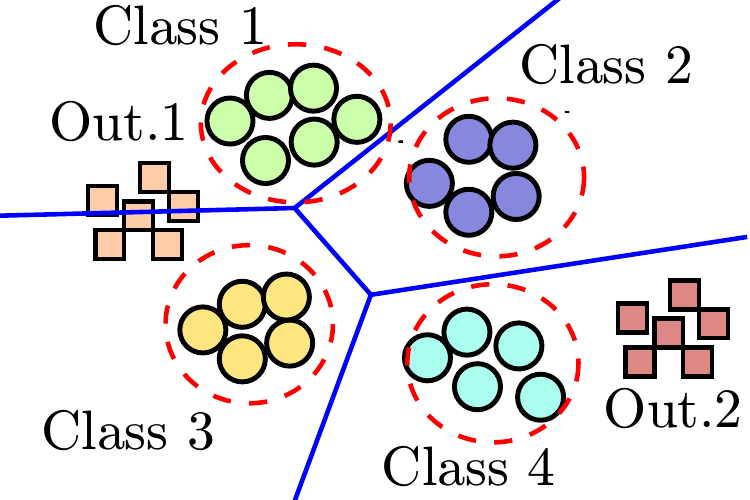}
	\caption{Known classes are depicted as circles and unknown classes as squares. Solid lines and dashed circles represent classification boundaries and open set recognition, respectively.}
	\label{fig:openset_anomaly_class}
\end{figure}

 In this paper, we pose the problem of rejecting signals from \text{any} transmitter outside a known authorized set as an open set recognition problem. Since this problem was studied extensively, instead of reinventing the wheel, we aim to adapt and evaluate well-established approaches for transmitter authorization. Since  the number of authorized transmitters is a system requirement that can vary significantly, we study how these approaches scale in terms of performance and network size with it. To further improve the performance, we propose using a set of known unauthorized transmitters and demonstrate its effectiveness in improving outlier detection. Our results show an average accuracy of 96.8\% when separating signals of 10 authorized transmitters and 30  unseen transmitters when using a set of 25 known outliers for training using a WiFi capture.
 The problem of open set recognition is more challenging than closed set classification. A closed set classifier  determines boundaries that separate the classes it has seen, as shown with the solid blue line in Fig.~\ref{fig:openset_anomaly_class}. But, given data from new classes (new unauthorized transmitters), the classifier will predict the nearest class, which poses a security risk for an authentication system. On the other hand, open set classification creates boundaries around the seen distribution, as illustrated with the red dashed circles in Fig.~\ref{fig:openset_anomaly_class}, for rejecting samples from new classes. 
  Unlike feature based approaches \cite{gulati_gmm_2013,xu_device_2016-1,xiao_using_2008,nguyen_device_2011} which use well separated features like CSI, our approach needs to learn the features that separate authorized transmitters from transmitters for which no training data is available. %

The remainder of the paper is organized as follows: we start by formulating the problem in Section \ref{sec:problem_formulation}. Section \ref{sec:ml_approach} discuses the considered machine learning approaches. The dataset, the architecture, and the results are presented in Section \ref{sec:results}. Section \ref{sec:conclusion} concludes the paper.

\providecommand{\mA}{\mathcal{A} }
\providecommand{\mK}{\mathcal{K} }
\providecommand{\mO}{\mathcal{O} }

\providecommand{\mAc}{\mathcal{|A|} }
\providecommand{\mKc}{\mathcal{|K|} }
\providecommand{\mOc}{\mathcal{|O|} }

\section{System Model and Problem Formulation}
\label{sec:problem_formulation}
\begin{figure}[t!]
	\centering
	\includegraphics[scale=0.7]{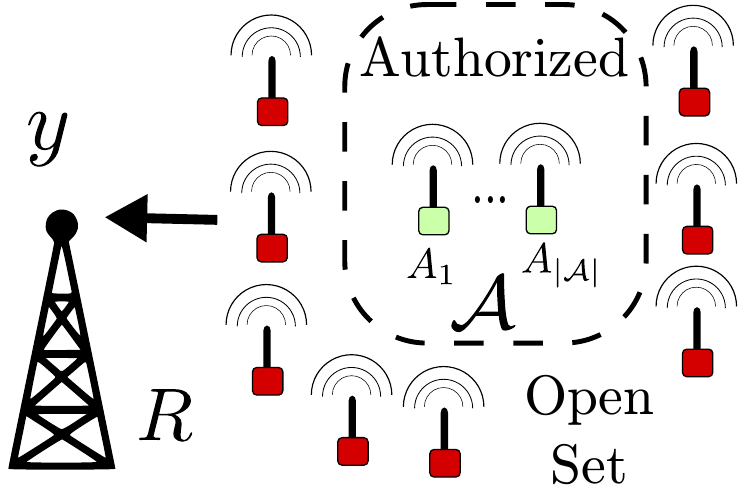}
	\caption{Signal $y$ is received by receiver $R$. We want to determine if it was sent by an authorized transmitter in the set $\mA$  or a new unseen transmitter.}
	\label{fig:problem_formulation}
\end{figure}

We consider a finite set of authorized transmitters given by  $\mA=\{A_1,A_2, \cdots,A_\mAc \}$ that are authorized to send data to a receiver~$R$, where $\mAc$ is the size of the set $\mA$. When a transmitter $T$ sends a set of symbols $x$, the signal received is $f_T(x)$. The function $f_T$ models the transmitter fingerprint determined by the variability of its circuit and also includes the effects of the  channel. The authorization problem can be formulated as shown in Fig.~\ref{fig:problem_formulation}: receiver $R$ receives a signal $y$ from some transmitter $T$ and wants to determine whether the transmitter $T$ belongs to the authorized set or not, based on $y$. This can be formulated as the following hypothesis testing: 
\providecommand{\mPfa}{P_{FA}}
\providecommand{\mPd}{P_{D}}
\newcommand{\mHz}{\ensuremath{\mathcal{H}_0} }
\newcommand{\mHo}{\ensuremath{\mathcal{H}_1} }
\renewcommand{\b}[1]{\boldsymbol{\mathrm{#1}}}
\begin{align}
\begin{split}
	\mHz: & \ y = f_T(x), T\in \mA \\
	\mHo: & \ y = f_T(x), T\notin \mA  
\end{split}
\end{align}  
Here, \mHz corresponds to an authorized transmitter and  \mHo corresponds to an outlier.

 Additionally, in cases where each authorized transmitter has different privileges, we might be interested in classifying the transmitter within the authorized set, which can be formulated as finding  $\hat{A}$ that maximizes the probability of identifying the true transmitter
\begin{equation}
\hat{A} = \underset{T}{\text{argmax}} \ \mathrm{Pr}(f_T(x)=y\ |\ y) , \ \ \ \ T\in \mA
\end{equation}

To improve the outlier detection,  we propose using an additional class of known outliers $\mK=\{K_1,K_2,\cdots,K_\mKc\}$, where $\mK \not\subset \mA$. Samples from transmitters in $\mK$ will be used during training to assist the outlier detector to differentiate between authorized and non-authorized transmitters. But still, the evaluation of any outlier detector is done using a set of unknown outliers  $\mO$ such that $\mO \cap \mK = \emptyset$. In practice, samples from the set $\mK$ can be obtained by capturing data from a finite number of non-authorized transmitters. 

\section{Machine Learning Approach}
\label{sec:ml_approach}
In this section, we  discuss the neural network architectures used to solve this problem and the processing performed on the output of these networks to decide if a signal is an outlier.
  We consider several neural network architectures for outlier detection. These networks consist of a feature extractor followed by one or many classifiers. In terms of training, some of these networks need known outliers  to generalize to unseen transmitters, while others don't. We also discuss how the size of $\mA$ affects the number of parameters of these networks.

\begin{figure}[t!]
	\centering
	\subfloat[Disc \label{fig:net_disc}]{\includegraphics{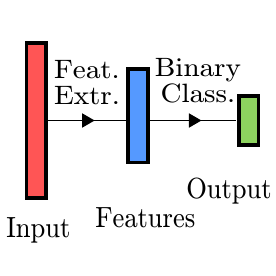}}  \hspace{0.5mm}
	\subfloat[DClass \label{fig:net_dclass}]{\includegraphics{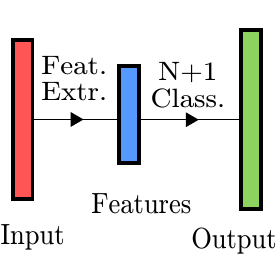}} \hspace{0.5mm}
	\subfloat[OvA \label{fig:net_ova}]{\includegraphics{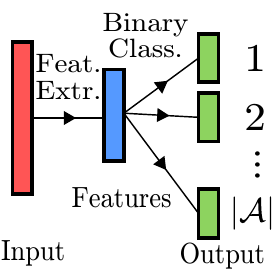}} 
	
	\caption{ Architecture of the proposed methods. }
	\label{fig:arch}
	\vspace{-5mm}
\end{figure}

\subsubsection{Discriminator (Disc)}
 One intuitive approach for outlier detection is to train a discriminator that outputs a decision on whether the signal is an outlier or not.  The  discriminator has a single scalar output $z$  as shown in Fig.~\ref{fig:net_disc}.  $z$ is generated by a sigmoid and takes a value between 0 and 1. The labels for authorized transmitters and outliers are $l=0$ and $l=1$, respectively. Samples with $l=1$  used in training are the known outliers, which are necessary for this approach.  In the test phase, we declare $\mHo$ if $z>\gamma$ for some threshold $\gamma$, else \mHz is declared.   In terms of architecture, Disc has the   advantage of having a fixed size regardless of $\mAc$.  Although this approach does not classify the authorized transmitters, a classifier can be cascaded with a discriminator to achieve this but is not discussed in this work.

\subsubsection{Discriminating Classifier (DClass)}
Instead of  cascading a discriminator and a classifier, we can directly train a network with $\mAc+1$ outputs, where the additional class corresponds to outliers.  Similar to Disc this approache relies on $\mK$ for training. As for deciding on outliers, a signal is classified as an outlier if the maximum activation corresponds to the  last class, else it is considered authorized without  adjustable thresholds.  In comparison with Disc, the labels of the transmitters are  expected to help DClass learn better features and hence perform better. This comes at the cost of increasing the size of the last layer   as $\mAc$ increases. 

\subsubsection{One Vs All (OvA)}
A simple way to  modify  Disc to include classification with modifiable thresholds  is to use $\mAc$ instances of it;  one  for each transmitter. However, this method  requires a high computational complexity due to having $\mAc$ feature extractors performing the same task. A better approach, proposed in \cite{shu_doc_2017}, is shown in Fig.~\ref{fig:net_ova}. In this approach, we use $\mAc$ binary classifiers, each deciding for a transmitter while sharing the same feature extractor.  The output of this network will be a vector $\b{z}$ of $\mAc$ real numbers such that $\b{0}\leq \b{z}\leq \b{1}$, where $\b{0}$ and $\b{1}$ are the vectors of all-zeros and all-ones, respectively. Following the notation in \cite{shu_doc_2017}, the labels for a sample from authorized transmitter $A_i$ will have   $l_i=1$ and  $l_j=0 ~\forall j\neq i$ while known outliers will have  labels equal to $\b{0}$.  For this architecture, the threshold will be a vector $\pmb{\gamma}$, where element $\gamma_i$ is the threshold for $z_i$. The binary classifier $i$ decides that the input sample belongs to class $i$ if $z_i>\gamma_i$; otherwise, it does not belong to class $i$. We declare the signal to be an outlier (corresponding to $\mHo$), if all discriminators declare the signal to be not within their class ($\b{z}\leq \pmb{\gamma}$), and to be within the authorized set (corresponding to $\mHz$) otherwise.

Note that OvA, unlike DClass and Disc, does not require a known set of outliers, since for samples of any class $i$,  $l_i=0$ for signals from other classes. OvA, however,  requires an entire binary classifier for each authorized transmitter. hence, among the proposed architectures, it has the worse scalability with respect to $\mAc$, in terms of the number of learnable parameters.

Both OvA and Disc have adjustable thresholds. The value of these thresholds determines their sensitivity to outliers.  A tight threshold would lead to signals from authorized transmitters being mistakenly rejected (high probability of false alarm $\mPfa$) and a loose threshold would fail to recognize many outlier signals (low probability of detection $\mPd$). This trade-off is commonly visualized by the receiver operating characteristic  (ROC) showing both $\mPfa$ and $\mPd$ for a specific receiver. 
 In the Appendix, we describe how this trade-off is implemented and state how a specific threshold is chosen to calculate the outlier detection accuracy.

\section{Experimental Evaluation}
\label{sec:results}
We start by describing the dataset used and evaluate the performance of the proposed network architectures on the dataset as we change the size and composition of $\mA$ and $\mK$.

 \subsection{Dataset}
 The dataset was captured using off-the-shelf WiFi modules (Atheros 5212, 9220, and 9280) as transmitters and a software defined radio (USRP N210) as a receiver, from the Orbit testbed \cite{orbit_2005}. The choice of the Orbit testbed is made due to the ease of access to many transmitters using realistic hardware while being able to isolate external environmental disturbances. The nodes in Orbit are organized in a 20$\times$20 grid with a separation of one meter; the receiver was chosen near the center and 71 transmitters  were randomly chosen, to make many nodes experience similar channels due to the symmetry.

The capture was done over Channel 11 which has a center frequency of 2462 MHz and a bandwidth of 20 MHz.  All transmitters were configured to have the same fake MAC address and same IP address to avoid providing any signal based clues about the identity of the transmitter. Captures were taken at a rate of 25 Msps for one second. After the IQ capture was complete, the packets were extracted using energy detection. The number of packets obtained from each transmitter during the capture period varied between 200 and 1500  packets with a mean of 800 packets. This variability is due to WiFi rate control. From each packet, we used the first 256 samples, containing the preamble, without any synchronization or further preprocessing.

\subsection{Network Architectures and Training}
\begin{figure}[t!]
	\centering
	\includegraphics[scale=0.78]{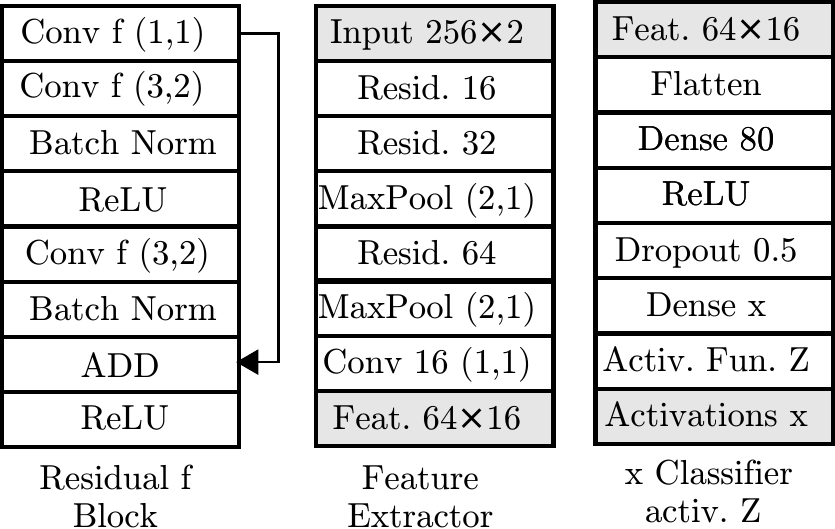}
	\caption{Detailed architecture of the feature extractor (made of residual blocks with f filters), and a classifier with x outputs.}
	\label{fig:net_layers}
\end{figure}
We consider the three previously proposed architectures Disc, DClass, and OvA. As stated earlier, these architectures consist of a feature extractor that processes the raw IQ samples and outputs features, followed by a number of classifier blocks. Our focus in this paper is on the approach, not the architecture, so, we used the same  feature extractor for all networks. The feature extractor consists of a series of residual blocks with different numbers of filters as shown in Fig.~\ref{fig:net_layers}. As for the classifier blocks, the architecture for each block is shown in Fig.~\ref{fig:net_layers}. This network was chosen because similar networks have shown superior performance to CNNs on a similar problem~\cite{oshea_over_2017}. For Disc, we used one classifier with a sigmoid activation. For OvA, we used $N$ of the classifier blocks with each block similar to that of Disc. As for DClass, we used one classifier block with $\mAc+1$ outputs and softmax activation.  L2 regularization was used in the dense layers with a weight of 0.001 to avoid overfitting. 

Note that for OvA and DClass the number of parameters of the neural network increases as the size of $\mAc$ increases. In OvA, for each new authorized transmitter, a new instance of the classifier is added to the architecture with about 80K parameters. For DClass, the size of the last layer increases by 81 parameters for each authorized transmitter. As for Disc, the number of parameters is constant for any $\mAc$.

The training was done for 10 epochs using the ADAM optimizer with a learning rate of 0.001. The weights with the lowest validation loss are kept. Samples was first normalized, then augmented by adding noise with a variance of 0.01 and applying a uniformly random phase shift. Cross-Entropy was used as the loss function with classes weighted depending on the number of samples for each class.

\subsection{Transmitter Set Sizes Evaluation}
Ideally, we want to train our network using the set of authorized nodes only, regardless of  their number. Creating a known outlier set $\mK$ would require more transmitters and more data collection.
In this section, we explore the effect of changing the size of the authorized set $\mAc$, and the size of the known outlier set  $\mKc$, on the ability of the network to distinguish authorized signals from outliers. We start by describing the evaluation metrics and dataset division.

\subsubsection{Evaluation Metrics and Dataset Division}
 Since certain subsets of the set of transmitters in our dataset might have more mutually similar signals than others, we try to make our results less specific to a chosen subset of transmitters. To this end, we randomly populate the sets $\mA$, $\mK$, and $\mO$ from the 71 transmitters 10 times in each test we conduct. Results are shown as mean and standards deviation of these 10 realizations. The metrics used for the evaluation of outlier detection are the accuracy and area under the ROC curve (AUC). The accuracy is the percentage of correct  predictions calculated over a balanced test set, such that any random or trivial guess would yield 50\% accuracy. %
  The area under the ROC curve provides a metric of which model is better on average \cite{sun_classification_2009}, while the accuracy is what we get for a specific threshold. Although DClass and OvA are capable of classifying signals within the authorized sets, the results of classification were above 99\% on the authorized part of the test set, and as classification has been extensively studied in the literature, we omit these results for brevity.

Our training, validation, and test sets are built as follows: for certain values of $\mAc$, $\mKc$, and $\mOc$,   we randomly choose transmitters to form our sets $\mA$, $\mK$, and $\mO$.    For training and validation, we use 70\% of the samples belonging to $\mA$, and all the samples belonging  $\mK$.  The shuffled combination of this data is split into 80\% for training and 20\% for validation. The test set contains all samples from $\mO$  and the remaining 30\% of $\mA$. For different realizations of the sets, the dataset can get highly imbalanced. To avoid degenerate solutions, where the network always predicts the class with the majority of samples, the training loss is weighted.

\subsubsection{Authorized set}
\begin{figure}[t!]
	\centering
	\subfloat[AUC \label{fig:n_auth_auc}] {\includegraphics{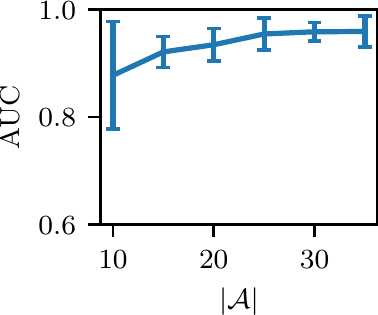}}
	\subfloat[Accuracy \label{fig:n_auth_acc}]{ \includegraphics{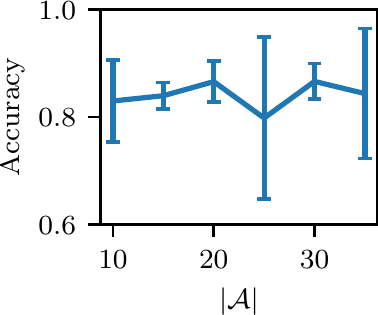}}
	\caption{Average performance of OvA as we change  $\mAc$. Error bars represent the standard deviation.}
	\label{fig:n_auth}
\end{figure}
We study how the size of the set $\mA$ affects outlier detection by considering OvA with no known outliers, $\mKc=0$.   Results are shown in Fig.~\ref{fig:n_auth} for $\mOc=30$; we see that as we increase the number of authorized nodes, the AUC increases and its standard deviation decreases.  In OvA, to generalize to unseen transmitters, binary classifier $i$ needs to learn the unique features of transmitter $i$.  Seeing signals from more transmitters helps it realize that leading to a better AUC. The accuracy is shown in Fig.~\ref{fig:n_auth_acc}, from which we can see that accuracy follows the same trend, except at the point with 25 and 40 authorized transmitters. At these points,  the chosen threshold for one realization resulted in a low accuracy, decreasing the mean and increasing the standard deviation. This shows that for specific combinations of authorized and outlier nodes, and a given threshold, we might get lower performance. Since the value of the AUC is high, this shows that another threshold would give a better performance. From these results, we can see that the average accuracy does not go over 90\%, while it fluctuates depending on the choice of transmitters for the same method of selecting thresholds.

\begin{figure}[t!]
	\centering
	\subfloat[AUC. \label{fig:n_kn_out_auc}] {\includegraphics{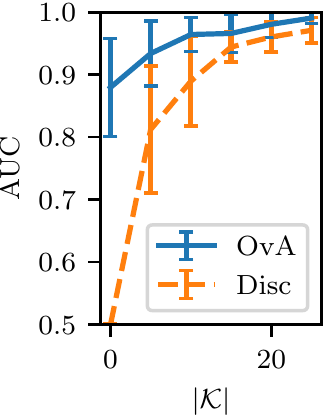}}
	\subfloat[Accuracy. \label{fig:n_kn_out_acc}]{ \includegraphics{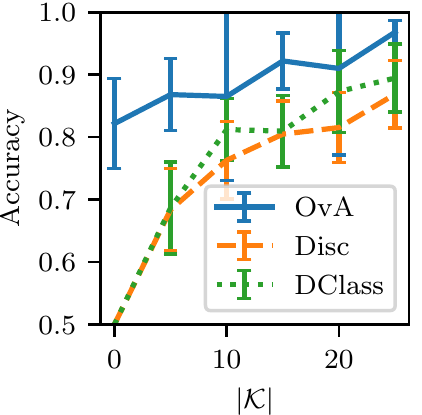}}
	\caption{Average performance of architectures as we change  $\mKc$ for $\mAc=10$. Error bars represent the standard deviation.}
	\label{fig:n_kn_out_auth}
	\vspace{-4mm}
\end{figure}

\subsubsection{Known set}
To further improve performance, we propose using known outliers to help the approaches generalize to unseen transmitters. We evaluate the performance of all architectures as a function of  $\mKc$ given $\mAc=10$ and  $\mOc=26$. The AUC and accuracy curves are shown in Fig.~\ref{fig:n_kn_out_auc} and ~\ref{fig:n_kn_out_acc} respectively. As stated earlier, at $\mKc=0$, DClass and Disc don't have any outlier samples and predict everything as authorized. From Fig.~\ref{fig:n_kn_out_auc}, we see that the performance of both OvA and Disc improves as we increase the number of known outliers. We note that OvA is performing better. This is explained by recognizing that in OvA each binary classifier sees more samples to reject; the known outliers and the samples not belonging to its class. Thus, it is able to isolate its class better. DClass and Disc, on the other hand, only learn to reject samples from $\mK$.
Fig~\ref{fig:n_kn_out_acc} follows the same trend with OvA reaching accuracies up to 96\% on average. DClass slightly outperforms Disc because  the labels of $\mA$ help it extract better features compared to Disc.
So even if we are not interested in classifying among the nodes in $\mA$, including these labels in training improves the outlier detection performance.
 In  the case with $\mKc=0$,  we were not able to attain accuracies above 90\% on average, showing that using additional unauthorized transmitters in data collection can lead to significant improvements.

\section{Conclusion}
\label{sec:conclusion}
In this paper, we formulated  transmitter authorization  as an open set classification problem where we learn to reject signals from new transmitters not seen during training. We  considered three approaches to solve it. OvA does not require known outliers and gives better performance, at the cost of a large increase in the neural network size with the size of the authorized set. DClass and Disc need a large set of known outliers for training to get good performance, with Disc having a slightly lower performance with the advantage of maintaining a constant network size regardless of the number of authorized nodes. In all cases, we have shown that having a set of known outliers improves performance.
So far, we have only considered a residual neural network. Further work needs to be carried out to understand the effect of changing the neural network type and architecture for open set recognition. %

\appendix
\section{Threshold Selection}
For OvA and Disc, we describe how  the threshold used for accuracy is calculated along with the ROC scan.
\subsubsection{Disc}   Ideally, we want the threshold to be as low as possible without falsely rejecting authorized transmitters. This can be done by adapting the threshold to tightly fit the predictions of authorized signals in the training set. We follow the approach proposed in \cite{shu_doc_2017}, where the predicted output  for the authorized (having labels equal to 0) is mirrored around 0 and fit to a Gaussian distribution having mean 0. Then, we calculate the standard deviation $\sigma$ of these samples and set the decision threshold to $\gamma= \min(0.5,  3\sigma)$.  As for obtaining the  ROC curve, we scan the value of $\gamma$ from 0 to 1. 
\subsubsection{One Vs All (OvA)}
To obtain the ROC curve, we scan a single threshold $\gamma$ from 0 to 1 such that $\pmb{\gamma}=\gamma\b{1}$. To calculate the accuracy, we use multiple thresholds designed according to the same method of  Gaussian fitting used in Disc.

\bibliography{references}
\bibliographystyle{ieeetr}

\end{document}